# The metaphysics of laws: dispositionalism vs. primitivism[1]

Mauro Dorato and Michael Esfeld
University of Rome III, Department of Philosophy, Communication and Media Studies
dorato@uniroma3.it
University of Lausanne, Department of Philosophy
Michael-Andreas.Esfeld@unil.ch

**Abstract**

The paper compares dispositionalism about laws of nature with primitivism. It argues that while the distinction between these two positions can be drawn in a clear-cut manner in classical mechanics, it is less clear in quantum mechanics, due to quantum non-locality. Nonetheless, the paper points out advantages for dispositionalism in comparison to primitivism also in the area of quantum mechanics, and of contemporary physics in general.

1.   Introduction

There are three main stances with respect to laws of nature in current philosophy of science: Humeanism, primitivism and dispositionalism.[2] Roughly speaking, according to *Humeanism*, the world is a mosaic of local matters of particular fact – such as the distribution of point-particles in a background spacetime –, and laws are the axioms of the description of this mosaic that achieve the best balance between simplicity and informativeness or empirical content (see e.g. Lewis 1994 as well as Cohen and Callender 2009 and Hall unpublished).

According to *primitivism*, over and above there being local matters of particular fact – such as an initial configuration of point-particles in a background spacetime –, there are in each physically possible world irreducible nomic facts instantiated by the world in question, according to which the corresponding laws hold in that world. The laws, *qua* instantiated in a world or by the world, fix the temporal development of an initial configuration of matter (in a deterministic manner if the law is deterministic, in a probabilistic manner if the law is probabilistic) (for primitivism see, notably, Carroll 1994 and Maudlin 2007). Laws are therefore not made true by locally instantiated properties or local matters of particular facts; on the contrary, such properties are discovered and determined by the laws that hold in a world.

---

[1] We thank an anonymous referee for penetrating comments on two previous versions of this paper.
[2] A fourth notable view about laws, defended in Cei and French (2010) and French (2014), will be introduced in the next section, when discussing primitivism.

According to *dispositionalism*, the local matters of particular fact – such as an initial configuration of point-particles in a background spacetime –, instantiate a property (or a plurality of properties) that fixes the behaviour of these local matters of particular fact, for example the temporal development of an initial configuration of particles (either in a deterministic or in a probabilistic manner, the property being a propensity in the latter case). This property thus is a disposition or a power, and the behaviour of the local matters of particular fact is its manifestation. This property grounds a law in the sense that the latter is made true by the former, so that a law describes how objects that instantiate the property in question behave or would behave under various circumstances (if the property is a propensity, it grounds a probabilistic law that describes how objects that instantiate the property in question behave or would behave under various circumstances; note that propensities are not probabilities, but are that what grounds probabilities) (see notably Bird 2007 and Suárez 2014 for propensities). As we will see, dispositionalism can be further divided into a realistic *vs.* an antirealistic position about laws; the latter has been defended in particular by Mumford (2004, 2005a, 2005b).

In this paper we will assume that the main dividing line runs between Humeanism on the one hand and primitivism as well as dispositionalism on the other. Humeanism has to accept the whole distribution of the local matters of particular fact as a primitive, since the laws, being the axioms of the description of that distribution that achieve the best balance between simplicity and empirical content, supervene only on that entire distribution. In a nutshell, thus, what the laws of nature are, is fixed only "at the end of the world". It is not the laws that determine the development of the world, but it is the development of the world, in the sense of its spatiotemporal arrangement, that determines what the laws are (see Beebee and Mele 2002, pp. 201-205). By contrast, primitivism and dispositionalism have only to accept the initial conditions of the world – such as an initial configuration of point-particles in a background spacetime – as a primitive. The initial conditions, plus the fact that (i) certain laws are instantiated in the world in question (primitivism about laws) or that (ii) the instantiation of certain properties (dispositions) is part and parcel of the initial conditions (dispositionalism), fix the further development of the world.

The reason for this divergence is that Humeanism eschews a commitment to objective *modality*, whereas both primitivism and dispositionalism subscribe to it. According to Humeanism, there is nothing about any proper part of the distribution of the local matters of particular fact in a world that fixes what is physically possible and what is not possible as regards the rest of the distribution of the local matters of particular fact in the world under consideration. The physical modality in question is not "in re", but belongs to the model or is a purely linguistic feature of nomic *statements*. According to both primitivism and dispositionalism, by contrast, there is something about a proper part of the distribution of the local matters of particular fact in a world that fixes what is physically possible and what is not possible in the world at issue, because either laws or a set of dispositional properties respectively are instantiated everywhere in the world.

Consequently, not only on primitivism, but also on dispositionalism, modality is not grounded in anything that is not itself modal. Thus, the dispositions that ground the laws according to dispositionalism are not themselves grounded in non-dispositional properties, but are basic properties. Their modal nature is therefore fundamental. In other words, both primitivism and dispositionalism are committed to a primitive modality. The difference between primitivism and dispositionalism is that the former position spells out primitive modality in terms of laws being primitive, whereas the latter position traces laws back to properties that display an ungrounded – and hence primitive – modality.

In this paper we will try to adjudicate the dispute between dispositionalism and primitivism by taking for granted that both parties believe in the existence of laws as well as in that of properties. The difference between these two positions lies in the fact that the dispositionalist regards laws as secondary to the properties, while the nomic primitivist considers properties to be ontologically secondary in a sense to be further specified below. For instance, by looking at the debate from a dynamical perspective, dispositionalism locates modal aspects in matters of particular fact by taking them to instantiate properties that are dispositions or powers and hence modal properties, while primitivism holds that it is the universal validity of laws in space and time that determines the temporal development of parts of the world or of the world itself by determining which properties exist.

After presenting primitivism and its possible formulations in the next section, we move on to discussing two case studies, by contrasting in each of them primitivism and dispositionalism (we take for granted that dispositional essentialism does not need a further presentation on our part – see Bird 2007). The first case study (section 3) is about laws in classical mechanics and is meant to illustrate the central feature of dispositionalism, namely grounding something that looks like a governing character of laws in properties that are localised in entities that there are in the world. The second case study then casts doubt on this straightforward picture by showing that properties that are supposed to ground the laws of our world as it is described by contemporary physics cannot be *local* properties, but have to be global and holistic. Such a non-locality seems to imply that the distinction between primitivism and dispositionalism becomes blurred, at least to the extent that primitivism is committed to the idea that what happens locally in region R in virtue of what properties are intantiated in R depends on what holds globally in the world, in virtue of the spatiotemporal universality of laws. We will therefore investigate whether this distinction can be upheld and if so, whether it provides a reason to prefer dispositionalism to primitivism (or the other way round) (section 4).

2.    *What does primitivism about laws mean?*

Let us first of all distinguish *conceptual* primitivism about laws from *ontological* primitivism. The former amounts to the claim that the notion of law cannot be analyzed or reduced in terms of counterfactuals, causation, regularity, explanatory or predictive power, and the like, since all of these notions presuppose it. The latter claims that laws exist in a primitive way, so that the existence of properties is grounded, supervenient or dependent on the existence of laws.

Here we are interested in spelling spell out what it means to claim that laws are ontological prior to properties (*ontic* nomic primitivism), since this problem seems to have been left in the background even by nomic ontic primitivists: "My analysis of law is no analysis at all. Rather I suggest we accept laws as fundamental entities in our ontology. Or, speaking at the conceptual level, the notion of law cannot be reduced to other more primitive notions" (Maudlin 2007, p.18). Since Maudlin's clarification here moves from ontic priority to conceptual priority, in order to understand the sense in which the existence of properties may be grounded in that of laws, it will be opportune to start our discussion by clarifying the main features of *conceptual* nomic privimitism.

The latter view, defended among others also by Carroll (2004) insists on the "conceptual centrality of the nomic": "If there were no laws, then there would be no causation, there would be no dispositions, there would be no true (nontrivial) counterfactual conditionals. By the same token, if there were no laws of nature, there would be no perception, no actions, no persistence. There wouldn't be any tables, no red things, no things of value, not even any physical object." (Carroll 2004, p. 10). In this sense, conceptual primitivism about laws implies that the notion of law is necessary (and sufficient) to explain (note, an epistemic notion) the notion of physical possibility and therefore to specify the set of models that are consistent with the law (Maudlin 2007, p.18). In a clear sense then, a law is more fundamental than the notion of model because different models can share the same law (think of different cosmological models sharing the same laws, namely Einstein field equations). In addition, once we have the notion of law, that of counterfactual can also be explained: if "all *A*s are *B*s" is fundamental, then the fact that "if *x* were an *A*, it would be a *B*" follows.[3] The latter notion of counterfactual, in its turn, would provide an analysis of causation (a counterfactual theory of causation), and of dispositions: if a certain stimulus were to be applied to a glass or to a flammable match, the glass and the match would manifest their dispositions to break and to catch fire respectively. Also the notion of property (say, "being charged"), according to nomic conceptual primitivism, cannot but be analyzed by using nomic concepts as primitives. For instance, what charge *is* (its causal role) and what it *does* (its behaviour) depends or is derivative (for short, is grounded) on the particular laws in which the property of charge figures: the Coulomb law defines the behaviour of electrostatic charges, the Lorentz law fixes the behaviour of a charged particle entering an electromagnetic field, while the motion of charges creating an electromagnetic field is governed by the relevant Maxwell equation (see Roberts 2008, p. 65). In a word, not only are natural properties discovered by finding out what the laws are (*epistemic* priority of laws), but their causal role is also fixed by the laws (*ontic* priority of laws).

We can now move on to clarify what ontic nomic primitivism amounts to. There are at least *two* senses in which laws can be ontically prior to dispositional properties, which we will discuss in turn: the first is spelled out in terms of supervenience, the second in terms of a structuralist viewpoint on laws (Cei and French, French 2014):[4]

1) According to a first way to spell out the failure of supervenience of laws on properties, "two worlds could differ in laws but not in any observable respect" (Maudlin 2007, p. 17). Suppose that "observable respect" is read as "observable properties". Two worlds could have different laws but could share all observable, non-quiddistic properties. Of course, one could block this failure of supervenience if one *defined* the notion of property as something that essentially plays a certain nomic role, so that a difference in laws would automatically imply a difference in properties. But since this move would beg the question against nomic ontic primitivism, the real issue at stake is which of the two positions, dispositionalism or nomic conceptual primitivism, is more suitable to perform an explanatory role with respect to the other notions that are typically associated with that of law (necessity, possibility, model, causation, counterfactual, regularity, etc). The next step would then be to ask whether such an explanatory role due to conceptual priority suffices for an inference to the best explanation *vis à vis* ontological priority. But also an inference to the best explanation in this case would be suspicious for both primitivism and dispositionalism. In fact, one could argue that even though the concept of law or of dispositional property are non-reducible to non-nomic concepts and are furthermore explanatory primary in being indispensable to analyse causation, dispositions and counterfactuals, etc., there is nothing in the world that corresponds to laws of nature (nomic antirealism about laws) or dispositional properties. In sum, it seems that antirealists about any type of modal notions may coherently recognize that laws or dispositional properties are primitive *only* on the conceptual level.

---

[3] For a contrary view, see Lange (2009).

[4] A third interesting analysis is provided by Lange (2009), who claims that metalaws like the relativity principle constrain laws, and the latter constrain the behaviour of physical systems and their properties.

We are convinced that stalemates of this kind in the metaphysics of science can best be settled by appeal to specific examples. Given the case study that we will discuss in the next section, for now it is appropriate to prepare the ground for the discussion by introducing a specific example, involving Einstein's and Newton's laws of gravity vis à vis the dispositional property "being massive". Suppose along with the ontic nomic primitivist that a difference of laws in a spacetime region $R$ did *not* require a difference in the properties $P$ instantiated in $R$. Since Einstein's and Newton's laws of gravity are *different*, it follows that the ontic primitivist must argue that this difference is compatible with the fact that the property of being massive in the two cases (or the two possible worlds in which these laws hold) stays *the same*. For instance, granting that "causing acceleration" is essential to "being massive", the nomic ontic primitivist might insist that the latter property has the same causal role in the two different laws. On the other hand, the reply of the dispositionalist might consist in pointing out that the mediating role in the manifestation of a dispositional property is essential for establishing the causal role played by a property. In Newton's case, the acceleration is mediated by a force, but in general relativity, masses accelerate via the mediation of the curvature of spacetime, and also in this case gravitational forces are non-existent. So the property "being massive" in the two worlds would be different. Given that we will discuss these two different readings of the property "being massive" in the next section, we can move on to the second way of cashing nomic ontic primitivism, which is in terms of the identity of properties.

2) The second way of cashing out the priority of laws over properties instantiated by local matters of facts, is suggested by the structuralist understanding of laws proposed in Cei and French (2010) and French (2014). Despite the fact that these authors do not interpret their position as a kind of *nomic* ontic primitivism, quotations as the following seem to authorize this interpretation: "objects, whatever their status might be, do not enter certain lawlike relations in virtue of certain ontological aspects of their properties; rather *their properties present certain ontological aspects because of the relations they enter*" (Cei and French 2010, p. 11). The relations in questions are the nomic structures, which can be regarded as epistemic or ontic, as in the case of the distinction between epistemic and ontic structural realism. Also in the nomic case in fact, we can have an *epistemic* structural primitivism about laws and an *ontic* structural primitivism about them.

The former insists on the fact that all we know about the world are nomic relations, and locally instantiated properties are discovered by discovering the laws.[5] This might be regarded as meaning that the nomic structure in question is fixed by certain spatiotemporal symmetries that the world instantiates (consider the conservation laws as they are explicated by Noether's theorem), and objects and their properties are discovered in terms of what is left invariant by these symmetries. The ontic priority of laws can be construed either in an eliminationist fashion (there is just nomic structure and no local matters of facts instanting natural properties) or as moderate form of the structural primitivity of laws: both laws and properties are real, but the latter are grounded in the former, whatever grounded may mean in this case.

The connection between this way of construing the priority of laws with respect to dispositions and the first just sketched in 1) is given by the fact that if the properties get their *identities* from the laws in which they occur (as in Ramsey-style versions of structuralism), then the properties' identities supervene on laws. Worlds with the same laws must have the same properties. Of course, the problem in this second version of ontic primitivism is to characterize the nomic *ontic* structure in a clear way, a problem that notoriously besets the ontic version of structural realism and that here cannot be discussed.

In a word, primitivism about laws of nature is the view that there are nomic facts holding in each possible worlds that determine or at least put a constraint on the distribution of the local matters of particular facts in each of the worlds , *

## 3. Dispositions and laws in classical mechanics

According to dispositionalism, it is in virtue of having a mass *m* that particles exert a force of attraction *F* upon each other as described by the law of gravitation:

$$F = G\frac{mm'}{r^2} \tag{1}$$

On dispositionalism therefore, this law tells us that if two masses change their velocity due to the action of forces on them, the generated forces can be traced back to or explained in terms of the properties of the particles.

In other words, mass is a disposition that manifests itself in the mutual attraction of massy objects. The presence of another mass *m'* acts as a stimulus on *m* (and conversely) for the manifestation of the disposition in terms of a mutual acceleration. As soon as there are at least two massive objects in a world, that disposition is triggered. It is essential for the property of gravitational mass to manifest itself in the mutual attraction of the objects that instantiate this property. That's what gravitational mass *is* – the property that makes objects accelerate in a certain manner. It is in this sense that the dispositional property "having a mass" grounds the law of gravitation. More precisely, mass as a property type grounds the law (1), with the concrete values of mass – the mass tokens – determining, together with the square of the distance between the massy objects, how these objects attract each other in virtue of possessing each a certain value of mass. Hence, that law reveals and describes what objects do in virtue of possessing a mass, and, crucially, in Newton's mechanics this property depends on its manifestation (namely, the acceleration) on the existence of a force. Since in Einstein's theory the notion of force is jettisoned, we cannot consider that the notion of mass in two theories is the same, since in the latter case the manifestation depends on the curvature of spacetime: therefore, different laws imply different properties, and laws supervene on the dispositional property mass.

---

[5] In this perspective, for example, Chakravartty's detecting *properties* (2007) would depend on the nomic structure, and not conversely.

We can now take up the questions mentioned in the first section: assume that the local matters of particular fact consist in the distribution of point-particles in a background spacetime. Given an initial configuration of particles, that configuration develops in time in such a way that the particles trace out certain trajectories in space according to the laws of classical mechanics. Dispositionalism about Newton's laws (the first one in particular) maintains that the particles have the disposition to continue to move with constant velocities on straight lines in space (or to continue to be at rest), *unless* external forces act on them. This dispositional property grounds Newton's first law, and also grounds, or is identical with, the tendency to resist acceleration (inertial mass). The possible non-existence of inertially moving systems (nothing can be screened off from gravitational forces) makes the posit of a disposition to continue with the same speed rather plausible or perhaps even indispensable. Such a disposition is in fact the *truth-maker* of Newton's first law, regarded as the statement found in textbooks and used for the construction of the mathematical model given by Newtonian spacetime. Positing instead a primitive nomic fact about inertially moving bodies (along with the primitivist) seems inappropriate, since the fact in question might be, and most probably is, *uninstantiated*. How can a primitivist justify her position with non-instantiated laws? Note that this is a major problem also for Humean regularists, since they rely on the existence of concrete regularities in order to justify the existence of patterns of local facts, even if one claims that Newton's laws are axioms that maximize simplicity and informativeness. In any case, the burden of proof is on the side of the Humean to show how an *uninstantiated* regularity can be part of the regularities in a given mosaic of local matters of particular fact that allows to simplify that mosaic while being informative about it. In this respect, dispositionalism seems to fare much better than its two rivals.

The primitivist may raise at least two objections against this argument.[6]

1) Requiring a truth-maker for laws begs the question against primitivism. After all, positing a truth-maker automatically implies that laws are not primitive, as they must be grounded in something else!

2) According to primitivism, laws determine the physically possible models. Models in their turn are often idealized representations of the properties of physical systems: in this way, primitivism can account for the fact that the first law might be uninstantiated (if indeed it is) as models are a limiting case of the behaviour of real systems.

The response to the first objection lies in two counterobjections. First, the truth-maker truth-bearer distinction is widely shared, especially among philosophers inclined toward scientific realism, and is therefore completely *neutral vis à vis* the debate we are interested in. Second, we need to disambiguate « law » in terms a distinction between laws of science (*statements* expressing scientific laws) and what they denote, namely laws of nature (Weinert 1995), which according to *both* camps exist independently of any such statements. We take it that these two distinctions are quite reasonable and should therefore by endorsed by both camps. But then, if one rejects the instrumentalist view according to which statements expressing scientific laws are neither true nor false, and if one endorses the first distinction, the fact that no *real* physical system obeys the first law has the consequence that any statement expressing it must be considered to be strictly speaking *false*. If regarding the first law of mechanics as false is the price to pay for primitivism, it must be admitted that dispositionalism in this account fares better. The second distinction reveals the *confusion* on which the first objection is based: it is not laws of nature that need grounding, but laws of science, and primitivism, unlike dispositionalism, cannot offer any grounding for statements expressing the fist law.

---

[6] We owe these objections to the anonymous referee.

The response to the second objection, related to the first, involves the notion of models as idealizations of real physical situations. As is well known, physical models are often regarded as *mediators* between the theory and the world (among others, see Morgan and Morrison 1999). Here we will assume that this is indeed the case also in our context. Then the question once again is : how can something that does not exist (an uninstantiated law of anture) ground abstract, idealized models by determining them? Such models would be grounded on *nothing* real. Furthermore, assuming that models are mediators between the theory and the world seems to imply that the idealizations of reality that feature in the model are fixed by our theories of the physical world and not by laws of nature. At least if theories are not deducible by the facts (in our case, primitive *nomic* facts) but are, as Einstein put it, «free inventions of the human mind». But while in dispositionalism the constraint on theories is given by real dispositions, in primitivism such a constraint would be rather weak to say the least, at least in the case of uninstantiated laws.

Of course, the fact that the first law might be non-instantiated (only a system that were completely removed from any gravitational mass would obey Newton's first law) does not imply that in classical mechanics there is no empirical distinction between an inertial and an accelerated system. On the contrary, the distinction in question cannot be explained by the primitivist, because the law in question, unlike the related disposition, does not or might not exist in the actual, concrete world.[7]

From this viewpoint, one could *prima facie* take *two* positions: antirealism and realism about laws. According to the first position, laws do not exist in nature, since dispositions do all the work that the latter are supposed to do (Mumford 2004, 2005a, 2005b) and laws of science are true descriptions of what dispositions do. As Ellis put it: "Laws are not superimposed on the world, but grounded in the natures of the various kinds of things that exist" (2006, p. 435). As such, they cannot govern at all, because they do not exist. Or, secondly, one can endorse *realism* about laws, but *analyze* it and ground it by using the existence of dispositions: that is, the fact that laws exist is tantamount to the fact that dispositions or relations among them manifest themselves in a certain way. As long as the dependence of laws on dispositions is clear, we think that it is not important to choose

---

[7] When it comes to classical general relativity, inertial motion is explained by the vanishing of the covariant divergence of the stress-energy tensor, but the geodesic principle can still be interpreted in a dispositionalist fashion and not only within the dynamical approach to relativity favoured by Brown (1995, pp. 160 ff.). We have no room to argue in favour of this claim here.

between these two positions: they both agree that laws are grounded in dispositions and then take different stances with respect to the ontological status of non-fundamental entities

However, already in this paradigmatic example of a disposition grounding a law, other complications arise. Considering the formula (1), it seems that one can hold the masses $m$ and $m'$ fixed, but conceive a possible world in which the gravitational constant $G$ has another value, or a world in which the force of gravitation does not decrease with the square of the distance $r^2$ among the particles, but only with the distance $r$, or with the cube of the distance $r^3$, etc. It seems that in all these possible worlds, there is mass as in the actual world, but the law of gravitation is different, although it still is a law of gravitation.

Both Humeanism and primitivism admit such a scenario. On Humeanism, it is a contingent matter of fact that in the actual world the property which we refer to by using the term "mass" plays a role that is described by the law of gravitation that holds in the actual world. The role that this property plays can vary from one possible world to another. On primitivism, it is a primitive matter of fact that the law (1) is instantiated in the actual world. In other logically possible worlds, different, but similar laws are instantiated, which can also be considered as laws of gravitation.

In order for the dispositionalist to maintain that the law (1) is grounded in the property of mass – so that, whenever in a possible world there are objects that instantiate the property of mass, the law (1) applies –, the dispositionalist has to hold that the property of mass includes not only what is represented by the variable $m$ in the formula (1), but also the gravitational constant having a certain value and the fact that the force of acceleration that objects exert upon each other in virtue of possessing a mass decreases with the square of the distance. In other words, the dispositionalist has to pack everything that the law of gravitation says about the interaction of massive objects into the property of mass, in such a way that this property can ground the law. Making this move has the following consequence: since according to dispositionalism the role that a property exercises is the essence of the property, the dispositionalist is committed to maintaining that in a possible world in which the gravitational constant has another value, or in which the force of gravitation does not decrease with the square of the distance $r^2$, the property of mass is not instantiated. In such other possible worlds, another property is instantiated which is similar to the property of mass that is instantiated in the actual world. However, it is not mass, but only its counterpart. There is mass if and only if the law of gravitation as expressed in formula (1) holds. Note that this is an ontological issue; we may of course be in error about the dispositional essence of mass and therefore misconceive the law of gravitation, or amend our conception of that law through theory change. These epistemological matters have no bearing on the fact that according to dispositionalism, there is mass if and only if a particular law holds, whether or not we are right about what that law is.

Making this move has a clear advantage: it grounds everything that there is in a world for the law of gravitation to hold in an intrinsic property of the objects that there are in the world, namely the massive particles. Hence, thanks to this move, dispositionalism is committed in this case, like Humeanism, only to local matters of particular fact, namely particles instantiating certain properties and not also to locally instantiated laws fixing those properties. What distinguishes dispositionalism from Humeanism is that dispositionalism conceives these properties as modal, so that a world in which the gravitational force decreases with the inverse cubic power of $r$ is only logically but not physically possible.

Let us briefly turn to another paradigmatic example from classical mechanics, namely charge. According to dispositionalism, in virtue of possessing a charge (positive or negative), particles exert a force of attraction or repulsion upon each other as described by the laws of electromagnetism (Lorentz's equation included). In the case of Coulomb's law for example, each particle acts as a stimulus for the manifestation of the disposition of the other one to be attracted or repelled, and therefore to accelerate: the electrostatic force in this respect is fully analogous to the gravitational force. The classical theory of electromagnetism, however, distinguishes itself from Newton's theory of gravitation in that the force that particles exert upon each other in virtue of possessing a certain property is mediated by a *field*, so that the effects of this property are typically retarded (but advanced solutions exist, given the time-symmetric character of Maxwell's equations). In the case of Lorentz's force law, for example, the magnetic field triggers the disposition of the charge to be accelerated by the field, and the deviation of its trajectory is the manifestation of the disposition, while the motion of charges in a circuit manifests itself as a change in the magnetic field, which is disposed to be changed by a current, etc. In a word, as in the case of the gravitational law, the typical fingerprint of dispositions is present: a trigger mechanism and the manifestation of the disposition, either of the charge or of the field.

In sum, despite the mentioned difficulties and despite many details that have to be filled in, the dispositionalist can make a case for the fundamental laws of *classical* physics being grounded in dispositions that are intrinsic and thus local properties of particles or regions of fields.As we have seen, a first advantage of dispositionalism over primitivism is rather evident in the case of non-instantiated laws, of which we discussed only the law of inertia. Furthermore, since we have seen that the property of mass is given by its causal power, the mediation of force in one case of the gravitational curvature on the other makes it the case that the property mass is different in Newton's and in Einstein's theory of gravity, so that laws supervene on properties. Finally, the dispositionalist might claim to take up the advantages of both Humeanism and primitivism, while avoiding the drawbacks of each of these positions: like the Humean, the dispositionalist is committed only to local matters of particular fact; however, since these local matters of particular fact instantiate modal properties in the guise of dispositions or powers (such as mass, charge and local regions of the electromagnetic field), these properties ground laws in the sense of primitivism about modality, namely laws that implement an objective and irreducible modality. Since simplicity, however, need not be a guide to truth, the final balance between the two camps vis a vis the last requirement must be drawn in the next section.

*4.    Dispositions and laws in quantum mechanics*

Let us now turn to quantum physics and focus on what is known as "primitive ontology" approaches. [8] These are approaches that admit an ontology of matter distributed in three-dimensional space or four-dimensional spacetime as the referent of the formalism of quantum mechanics and propose a law for the temporal development of this distribution of matter. The motivation for doing so is to obtain an ontology that can account for the existence of measurement outcomes – and, in general, the existence of the macroscopic objects with which we are familiar before doing science. Here we will focus on three different primitive ontology approaches that have been developed in the philosophical literature on non-relativistic quantum mechanics (see notably Allori et al. 2008).

---

[8] Here the term "primitive" is confusing, but clearly there are two senses of "primitive" in play, one referring to what exists concretely in spacetime, the second to laws as being conceived as conceptually and ontically prior to properties or dispositions.

There is in the first place Bohmian mechanics, which is committed to an ontology of particles. This theory conceives a law, known as the guiding equation, that employs the quantum mechanical wave-function in such a way that, in brief, the temporal development of the wave-function according to the Schrödinger equation supplies the temporal development of the configuration of particles in three-dimensional space, by yielding a velocity field along which the particles move (see the papers in Dürr, Goldstein and Zanghì 2013 for the dominant contemporary version of the theory going back to de Broglie 1928 and Bohm 1952).

There is furthermore the amendment of the Schrödinger equation proposed by Ghirardi, Rimini and Weber (1986) (GRW). The GRW equation has the purpose to modify non-linearly the Schrödinger equation in such a way that it can describe the temporal development of matter that is localized in three-dimensional space. As regards matter, there are two different proposals for a primitive ontology of matter in physical space put forward in the literature that use the GRW equation: according to the proposal set out by Ghirardi himself (Ghirardi, Grassi and Benatti 1995), matter is "gunky", there being a continuous distribution of matter in space, namely a matter density field. That field can contract spontaneously in order to form well-localized macroscopic objects (where the stuff is more dense), as described by the spontaneous localization (the "collapse") of the wave-function in configuration space. Bell (1987) took up the GRW modification of the Schrödinger equation in another manner, proposing an ontology of events in spacetime, which in today's literature are known as flashes (that term goes back to Tumulka 2006, p. 826). According to this ontology, there is an event (a flash) in four-dimensional spacetime whenever the wave-function in configuration space spontaneously localizes ("collapses") as described by the GRW equation. Consequently, these events are sparsely distributed in spacetime, there being no continuous sequences of events. Nonetheless, the distribution of these events can be quite dense in certain regions of spacetime, so that well-localized macroscopic objects are accounted for also in the flash ontology.

The structure of all these proposals is such that (i) an ontology of matter in space or spacetime is admitted as the referent of the quantum formalism and (ii) a law is proposed that describes the temporal development of the configuration of matter in physical space. The universal wave-function – that is, the wave-function of the whole configuration of matter in physical space – is nomological in the sense that it is part of the law of the development of the primitive ontology, by contrast to being a *concrete* physical entity on a par with the primitive ontology (see notably Dürr, Goldstein and Zanghì 2013, ch. 12, in the context of Bohmian mechanics). The reason is that the wave-function could not be an entity that exists in three-dimensional or four-dimensional spacetime: it could not be a field in physical space, since it does not assign values to spacetime points. If it is a field, it could only be a field on the very high-dimensional configuration space of the universe (if there are $N$ particles, the dimension of the corresponding configuration space is $3N$).

While it is an option to regard the universal wave-function as a field and thus as a physical entity existing in configuration space, this option is not plausible within the primitive ontology approach: it is unclear to say the least how the universal wave-function could perform the task that it has according to the primitive ontology approach to quantum physics, namely to determine the temporal development of the primitive ontology, if it were a physical entity on a par with the primitive ontology, but existing in another space. How could a field on a very high-dimensional space make matter move in three-dimensional space or four-dimensional spacetime? It seems that anything doing so has to be situated in the same space as the matter whose motion it determines. Furthermore, according to configuration space realism, the high-dimensional configuration space of the universe is fundamental, being the space in which the physical reality, namely the wave-function, plays itself out and evolves (see notably Albert 1996 and 2013). On the primitive ontology approach, by contrast, three-dimensional space or four-dimensional spacetime is the domain in which the physical reality is situated. Everything else that is admitted in this approach then is introduced through the role that it plays in the law that describes the physical reality in three-dimensional space or four-dimensional spacetime. It is therefore well motivated to regard the universal wave-function as nomological in the primitive ontology approach to quantum physics, by contrast to being a physical entity on a par with the primitive ontology, but existing in another space.

When it comes to spelling out what it means that the universal wave-function is nomological, the three general stances on laws mentioned above are available and defended in the literature: on primitivism, a law is instantiated in the world over and above the primitive ontology, incorporating the universal wave-function or the quantum state (see Maudlin 2007, in particular ch. 2). On dispositionalism, as we shall elaborate on below, the configuration of matter in physical space instantiates at any time a holistic property that grounds the law of motion and that is *represented* by the universal wave-function, as the mass or the charge variable in the laws of classical mechanics represent dispositional properties of the particles. On Humeanism, the universal wave-function is nothing in addition to the distribution of the primitive ontology (the particles, the matter density field, the flashes) throughout the whole of space-time; it supervenes on that distribution, figuring in the Humean best system, that is, the system that achieves the best balance between being simple and being informative in describing the distribution of the primitive ontology throught the whole space-time (see E. Miller 2013, Esfeld 2014, Callender unpublished).[9]

Although there is a good reason to regard the wave-function as nomological in contrast to being a physical entity on a par with the primitive ontology, one has to bear in mind the following two facts: at least as a *law of science* as it is formulated in the model, the universal wave-function develops itself in time according to a law, namely the Schrödinger equation (or the GRW equation), which for realists about laws does refer to a *law of nature* – unless one assumes that the universal wave-function will eventually turn out to be stationary, for instance in a quantum theory of gravitation that replaces the Schrödinger equation with the Wheeler-deWitt equation. But even if the universal wave-function were stationary, there would still remain the fact that there are many different universal wave-functions possible all of which fit into the same law (the Schrödinger equation, the GRW equation, or the Wheeler-deWitt equation). We will show below how dispositionalism is in a better position to accommodate these facts than primitivism.

Let us first point out two important differences between classical and quantum dispositions. The first one is that both in the Bohmian picture and (even more) in the dynamical reduction models, the dispositions of the quantum objects manifest themselves in a *spontaneous* manner, that is, they are not triggered by anything external. One may object that the lack of a clearly identified stimulus for the manifestation of a disposition makes the property in question non-dispositional.[10] However, this charge is unjustified, especially for those dispositions that are also propensities and are indeterministic in nature. The dispositional property to decay possessed by radioactive material manifests itself spontaneously, since the time at which the manifestation of the disposition occurs is utterly indeterminate, and uncaused by anything external. For the Bohmian, deterministic case, one should simply note that, given non-locality, there cannot be any *external* triggering mechanism for the manifestation of the particles' disposition to fix their velocity field since – as we are about to see in the next paragraph – the disposition in question is a *holistic* property instantiated by the whole particle configuration. Furthermore, why should spontaneous dispositions not qualify as such? If the dispute here is not purely semantic, disqualifying spontaneous dispositions seems question-begging. Consider David Miller's example of the disposition or propensity of today's world "to develop in a year's time into a world in which I am still alive" (Miller 1994, p. 189). This disposition obviously does not require an external stimulus to be manifested, because it is a global one too.

---

[9] As regards the ontology of the wave-function, see the essays in Albert and Ney (2013). Unfortunately, this book ignores the Humean supervenience view of the wave-function, even though it contains papers that, like the Humean, are antirealist about the wave-function (notably French 2013, Monton 2013).

The second difference has to do with the main feature of the quantum mechanical wave-function – the feature that marks the distinction between quantum and classical mechanics – which is its entanglement. That is to say, whenever one considers a configuration of matter that comprises more than one particle, it is in general not possible to attribute to each particle a wave-function that, when put into the dynamical law, correctly describes its temporal development (for an interesting attempt to defend the contrary view, see Norsen 2010). Only the universal wave-function, that is, the wave-function of the *whole configuration* does so. The entanglement of the wave-function accounts for quantum non-locality: the temporal development of any part of the configuration of matter in physical space depends on all the other parts (although, as shown by the decoherence of the wave-function in configuration space, that dependence is in many cases negligible for all practical purposes).

Despite these differences, dispositionalism can be applied to quantum mechanics in the same way as in classical mechanics. The quantum law that describes the temporal development of matter in physical space (the Bohmian guiding equation, the GRW equation) is grounded in a property of matter that is a disposition, manifesting itself in the way in which the distribution of matter in space develops in time. The main difference between classical and quantum mechanics is that in the latter, the law can only be grounded in a property of the configuration of matter as a whole, that is, a global and holistic by contrast to a local property.

Thus, on dispositionalism applied to Bohmian mechanics, the configuration of all the particles in the universe at any given time *t* (recall that we are presupposing Newtonian spacetime) instantiates a dispositional property that manifests itself in the velocity of each particle at *t*; *the universal wave-function at t represents that property* and, within the guiding equation, the wave-function expresses how that property manifests itself in the temporal development of the position of the particles (that is, their velocity) (see Esfeld et al. 2013, sections 4-5).

---

[10] We owe this objection to Steven French and Juha Saatsi.

On dispositionalism applied to the GRW quantum theory, on the matter density version, the matter density as a whole instantiates a dispositional property (more precisely, a propensity) that manifests itself in the temporal development of the matter density – notably in its spontaneous concentration around certain points in space – and that is represented by the universal wave-function and the probabilities for the temporal development of the matter density that the universal wave-function yields if it is plugged into the GRW equation. On the flash version of GRW, the configuration of flashes as a whole in its turn instantiates a dispositional property (more precisely, a propensity) that manifests itself in the occurrence of further, later flashes and that is represented by the universal wave-function and the probabilities for the occurrence of further flashes that the universal wave-function yields once it is plugged into the GRW equation (see Dorato and Esfeld 2010 for dispositions in GRW and grounding the GRW probabilities in propensities).

Hence, when it comes to quantum mechanics, dispositionalism loses a characteristic feature by which it distinguishes itself from primitivism about laws as far as classical mechanics is concerned: it is no longer possible to maintain that the laws are grounded in local or intrinsic properties of particles. If the dynamical law of quantum mechanics is grounded in a property of matter, that property can only be a global or holistic property of the configuration of matter as a whole. Dispositionalism thereby comes close to primitivism in the following respect: on primitivism, each possible world instantiates a global fact – a world-fact so to speak – that a certain dynamical law holds in the world in question. In a nutshell, quantum mechanics compels dispositionalism to join primitivism in going global, at least to the effect that primitivism, by relying on the nomic structural realism presented in section 2, is best formulated as a view that stresses the existence of universal spatiotemporal symmetries as the bedrock for the primitive existence of laws, and therefore for the supervenient existence of properties (Cei and French 2010).

However, by contrast to primitivism, dispositionalism has no problem in accommodating the fact that the quantum mechanical wave-function develops itself in time as Schrödinger's equation prescribes (whereas fundamental dispositional properties in classical physics – such as mass and charge – do not develop in time, a particle always possesses the same values of mass and charge). The temporal development of the wave-function tracks or describes in the mathematical model the temporal development of the dispositional property that the configuration of matter as a whole instantiates at a time. Thus, this disposition manifests itself not only in a certain temporal development of the configuration of physical entities that instantiate this property (a development described by the guiding equation in Bohmian mechanics), but in inducing or causing such a temporal development, this property also causes its own temporal development (described by the Schrödinger equation in Bohmian mechanics) (in the GRW theory, the GRW equation incorporates both these developments).[11] In general, a dispositional property can change in time without the law in which the property in question figures being subject to a temporal development. By contrast, a law-fact instantiated in the world is not supposed to change in time, and it is difficult to see how primitivism about laws could accommodate the difference between a universal wave-function changing in time and the law in which it figures not changing in time.

---

[11] Recall that the spacetime presupposed by these non-relativistic theories admits a privileged foliation, i.e., simultaneity is absolute.

Furthermore, dispositionalism has no problem in accommodating the fact that there are many different universal wave-functions possible, all of which fit into the same law. Consider, for example, two possible worlds described by Bohmian mechanics and assume that in these worlds there is the same initial particle configuration, but different initial wave-functions applying to the initial configurations, leading hence to different trajectories of the particles in these two possible worlds. Dispositionalism accounts for this case by maintaining that there are different values of the holistic, dispositional property of the particle configuration instantiated in these two worlds, so that these two worlds differ in the initial quantum state represented by the universal wave-function. But there is no nomological difference between two such worlds. By the same token, there can be the same initial distribution of particle positions in two possible worlds of classical mechanics and different distributions of mass or charge, leading to different trajectories of the particles. Hence, by grounding the law in a dispositional property of the particles – be it a local property, be it a global, holistic one – dispositionalism can admit different values that this property can take without these differences amounting to any nomological difference. This fact points to the failure of supervenience of properties on laws; since different properties do not entail different laws, the dependence of laws on properties invoked by primitivists in this case fails. By contrast, on primitivism, different universal wave-functions amount to a difference in the law facts instantiated in the worlds in question. In a nutshell, primitivism, in these quantum examples in particular, faces a dilemma: either it has to bite the bullet of conceiving the law as developing itself in time and as including differences that correspond to different initial wave-functions, or it has to conceive the universal wave-function as a physical entity.

In sum, comparing quantum mechanics to classical mechanics, the distinction between dispositionalism and primitivism about laws of nature is much less sharp in the former than in the latter: due to the entanglement of the wave-function, it is no longer possible in quantum mechanics to ground laws in local or intrinsic properties of particles. Nonetheless, laws can still be grounded in properties, albeit global ones (so the term "intrinsic" does not really apply), and doing so can still be regarded as an argument in favour of dispositionalism: this position can make intelligible how laws can "govern" the behaviour of objects – they are our epistemic access to what can causally influence the behaviour of objects in a clear and straightforward way, namely certain properties of objects. In short, it is the essence of the properties that objects instantiate to influence their behaviour in a certain manner: this claim of dispositionalist essentialism holds independently of whether the properties are local, being instantiated by the physical objects taken individually, or whether they are global, being instantiated by a configuration of objects as a whole. By contrast, it is unclear how the fact of certain laws being instantiated in a world could influence the behaviour of the objects in the world in question. The Humean objection against the governing conception of laws of nature hits primitivism, but it does not apply to dispositionalism, at least if it is legitimate to assume a primitive modality. We take it to be the decisive advantage of dispositionalism over primitivism to make the governing conception of laws of nature intelligible by anchoring the laws in the properties of physical objects, which also allows dispositionalism to maintain that there can be different initial values of these properties and that they can develop in time, without these variations touching the laws that the properties in question ground.